\documentclass[twoside,twocolumn,9pt]{article}
\usepackage{authblk}
\usepackage[super,sort&compress,comma]{natbib}
\usepackage[version=3]{mhchem}
\usepackage[left=1.5cm, right=1.5cm, top=2.0cm, bottom=2.0cm]{geometry}
\usepackage{balance}
\usepackage{times,mathptmx}
\usepackage{sectsty}
\usepackage{graphicx}
\usepackage{lastpage}
\usepackage{float}
\usepackage{fancyhdr}
\usepackage{fnpos}
\usepackage[english]{babel}
\usepackage{array}
\usepackage{droidsans}
\usepackage{charter}
\usepackage[T1]{fontenc}
\usepackage[usenames,dvipsnames]{xcolor}
\usepackage{setspace}
\usepackage[compact]{titlesec}
\usepackage{hyperref}
\usepackage{subfigure}
\usepackage{amssymb}
\usepackage{caption}
\usepackage{xfrac}

\begin{document}
\title{Fusing machine learning strategy with density functional theory to hasten the discovery of MXenes for hydrogen generation}
\author{B. Moses Abraham$^{1\dagger}$, Priyanka Sinha$^{1\dagger}$, Prosun Halder$^1$ and Jayant K. Singh$^{1,2\ast}$}
\affil{\textit{$^1$Department of Chemical Engineering, Indian Institute of Technology Kanpur, Kanpur-208016, India. \\
$^2$Prescience Insilico Private Limited, Bangalore 560049, India.\\
$^{\ast}$E-mail: jayantks@iitk.ac.in}}

\date{}
\maketitle

{\Large\textbf{ABSTRACT:}}

The complexity of the topological and combinatorial configuration space of MXenes can give rise to gigantic design challenges that cannot be addressed through traditional experimental or routine theoretical approaches.  To this end, we establish a robust and more broadly applicable multistep workflow from the toolbox of supervised machine learning (ML) algorithms for predicting the hydrogen evolution reaction (HER) activity over 4,500 MM$^{\prime}$XT$_2$-type MXenes, where 25\% of the material space (1125 systems) is randomly selected to evaluate the HER performance using density functional theory (DFT) calculations. As the most desirable ML model, the random forest regression method with recursive feature elimination  and hyperparameter optimization  accurately and rapidly predicts the Gibbs free energy of hydrogen adsorption ($\Delta$G$_{H}$) with a low predictive mean absolute error  of 0.374 eV. Based on these observations, the H-atom adsorbed directly on top of the outermost metal atomic layer of the MM$^{\prime}$XT$_2$-type MXenes (site-2) with Nb, V, Mo, Cr and Ti metals composed of carbon based O-functionalization are discovered to be highly stable and active catalysts, surpassing that of commercially available platinum based counterparts. Overall, the physically meaningful predictions and insights of the developed ML/DFT-based multistep workflow will open new avenues for accelerated screening, rational design and discovery of potential HER catalysts. \vspace{0.05cm} \\

\textbf{KEYWORDS:} Transition metal carbides/nitrides, surface hydroxylation mechanism, electrocatalysis,  CO$_2$ conversion, first-principles calculations

\section{INTRODUCTION}

Growing concerns about the environmental problems and energy crisis demand the urgent development of affordable and clean renewable energy sources as a viable replacement for derogating fossil fuels. In this regard, electrochemical water-splitting is an effective and sustainable approach to generate a massive impact in clean-energy technologies\cite{RogerI,LewisNS,WalterMG}. However, the currently used expensive platinum group metals (PGMs) limit their large-scale applications, thereby promoting continuous research attempts toward highly active and non-noble metal electrocatalysts. Several promising candidates with zero or reduced content of PGMs are being considered, such as transition
metals\cite{McKoneJ} and their dichalcogenides\cite{YangJ,Voiry,ChenZ,XieJZ}, phosphides\cite{FengY}, nitrides\cite{CaoBF}, borides\cite{VrubelH}, carbides\cite{ChenWF} and metal-free carbon nitrides\cite{MerletC,MengSL}. Although massive experimental and theoretical studies demonstrate the usage of catalysts in hydrogen evolution reaction (HER), but the overall catalytic activity for large-scale hydrogen production is still confined to few active sites and poor electrical transport\cite{QTangD}. Therefore, it is of paramount significance to develop a broad range of catalytic materials with more active sites and higher conductivity, for which the fundamental understanding from an atomic scale point of view is highly essential.

\begin{figure*}[t]
\centering
{{\includegraphics[height = 3.5in,width=7.5in]{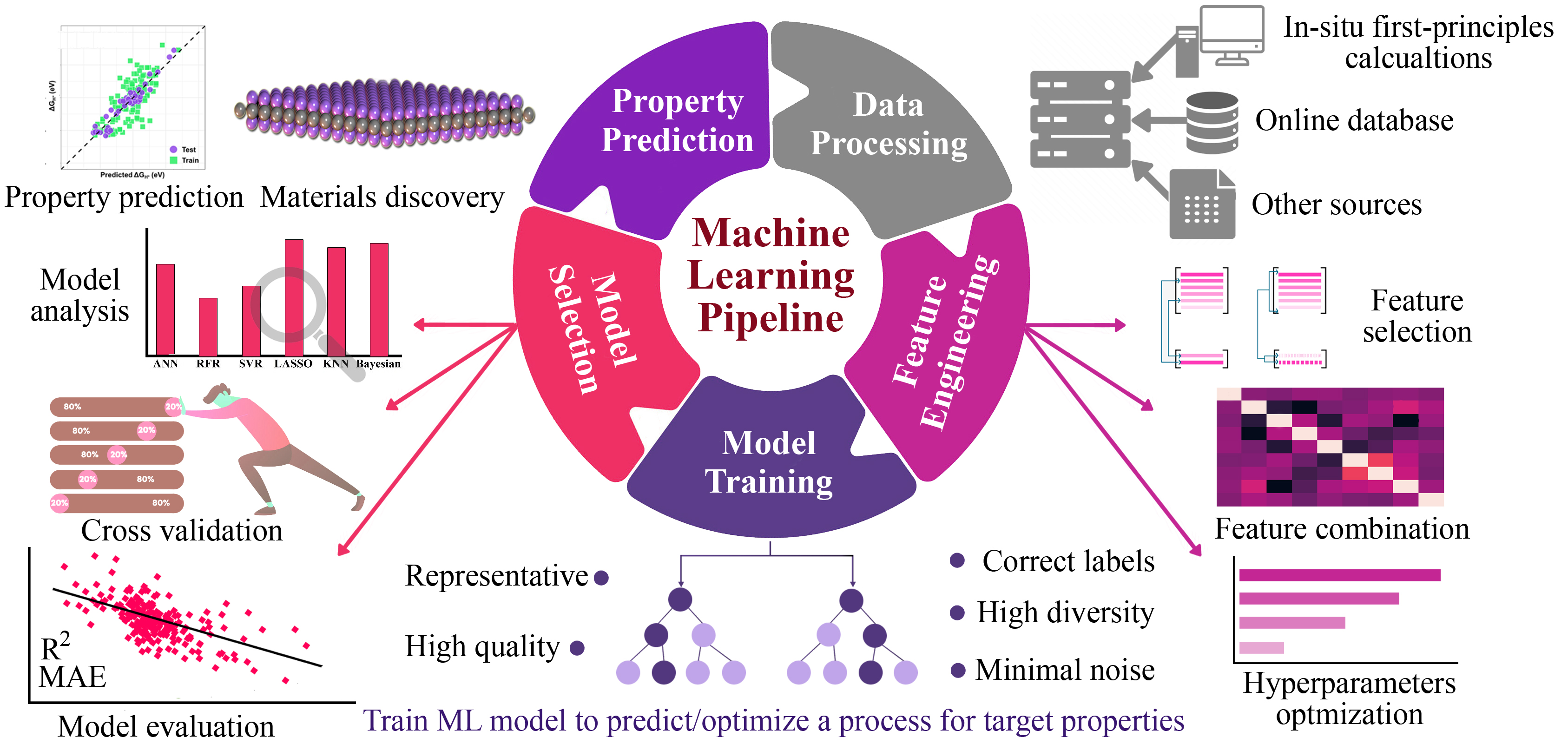}}}
\caption{Workflow of Machine learning approach, starting from data processing, feature engineering, model training, model selection and property prediction for screening of ideal HER catalyst from MXenes. From first principles calculations, materials' space is generated from a large number of possible combinations between selected elements and functionalization.}
\end{figure*}

MXenes, unique accordion-like structures exfoliated from MAX phases (M = transition metal; A = $\emph{p}$-block element; C = C and/or N), have recently attracted significant attention in electronic devices\cite{AChandra,MKhazaeiA,CSiK,MKhazae}, electromagnetic shielding\cite{FShahzad}, electrocatalysis\cite{JRan,ZLi,ZWShe,JZan}, energy storage and conversion\cite{XTangX,MRLukat,MNagui,JZho,Vanshree1,Vanshree2} applications. Especially, the long-term structural stability in acidic electrolytes\cite{PLiJ}, large active surface area (21 m$^2$/g)\cite{BWangAZ} and high electrical conductivity (4600 $\pm$ 1100 S/cm)\cite{ALipatov} make them suitable candidates for HER catalysis. In MXenes (M$_{n+1}$X$_n$T$_x$; n=1, 2, 3), tuning of M (transition metal), X (C and/or N) and T (surface functionalization) is found to improve the hydrogen evolution activity\cite{XHuiX}. For instance, manipulation of transition metal atoms in  M$_{n+1}$X$_n$T$_x$ (M$_{n+1}$X$_n$O$_2$, M$_{2}$M$^{\prime}$X$_2$O$_2$, and M$_{2}$M$^{\prime}_2$X$_3$O$_2$) leads to the identification of 110 unexplored candidates with better HER performance\cite{KuangP}. Sun et al.\cite{XSunJ} screened 271 different configurations of M$_{n+1}$X$_n$ by tuning X from C to B and found that the Mn/Co$_2$B$_2$, Os/Co$_2$B$_2$, Co$_2$B$_2$, Pt/Ni$_2$B$_2$ and Co/Ni$_2$B$_2$ candidates surpass the HER activity of PGMs. Doping of P region elements (surface functionalization) modulates the in-plane surface atom activity and improves the HER performance, thereby leading to an optimal HER Gibbs free energy\cite{YYoonA}. MXenes can also be used as substrates in HER applications because of their adjustable surface structures as well as promising physicochemical properties.\cite{GaoGO,ChengYZ}. In such cases, the performance of Ti$_2$CO$_2$ at various hydrogen coverages is found to improve by doping of S atom to substitute the surficial O atom\cite{WangSC}. NiS$_2$@VMXene exhibits long-term durability and low HER overpotential\cite{KuangP}. The aforementioned configurational space offered by the broad range of MXenes and their active sites using traditional approach for optimization of catalyst via experimental and theoretical screening is particularly challenging, time-consuming and expensive. Thus, finding suitable advanced methods became an essential task for accelerating the rational design of efficient catalysts.

The screening of potential MXene based catalysts from tremendous combinatorial and structural space requires a huge amount of computational resources\cite{Mosesab}. In traditional routine simulations, the H-adsorption energy is typically the most important parameter to evaluate the HER activity\cite{Marti}. According to the Sabatier principle, the binding of hydrogen should be neither too strong nor too weak to obtain the best catalytic activity\cite{Greeley}. However, such direct simulations might not provide complete information regarding HER performance since the descriptors in various reaction processes are equally important. In this regard, the incorporation of physical interaction through scientific knowledge into models trained by data-driven approaches has gradually emerged as a powerful and reliable tool for hastening the identification of catalysts\cite{Oriol,Mazheik,Emanuele}. Especially, random forest regression (RFR), support vector regression (SVR), kernel ridge regression (KRR) and Elman Artificial Neural Networks (Elman ANNs) algorithms are typically employed to predict Gibbs free energy, which is widely accepted descriptor of HER activity. For instance, the regularized random forest learning method reveals Ni-Ni bond length as the primary feature in determining the binding strength of hydrogen on Ni$_2$P (0001) plane\cite{RBWexler}. Sun et al.\cite{MSun} predicted the HER performance of graphdyine based atomic catalysts using the bag-tree learning model. These results demonstrate that the ML models not only discover novel catalyst materials, but also empower an in-depth understanding of the fundamental correlation between the catalytic structures and their properties. This is highly essential to modify the strategies for developing new design principles in revamping the electrocatalytic efficiency.

Here, we explore a robust and more broadly applicable multistep workflow as shown in Fig. 1, where the ab initio adsorption properties are combined with supervised toolbox of machine learning algorithms for source, verification and predictions. For this purpose, a data set of 4,500 MM$^{\prime}$XT$_2$-type MXenes was constructed and systematically investigated their HER performance. Among them, 1,125 systems (25\% of the materials' space) were randomly selected to evaluate the HER activity using DFT calculations as well as for training the ML model. Predominating indicators were then employed to build an interpretable ML model that predicts the HER performance of the remaining 85\% materials' space. Overall, the ML model achieves better prediction activity on par with the first-principles calculations. It deciphers the underlying factors that govern the HER performance and enables a coherent path to investigate a large

\begin{figure}[H]
\centering
{{\includegraphics[height = 7.9in,width=3.5in]{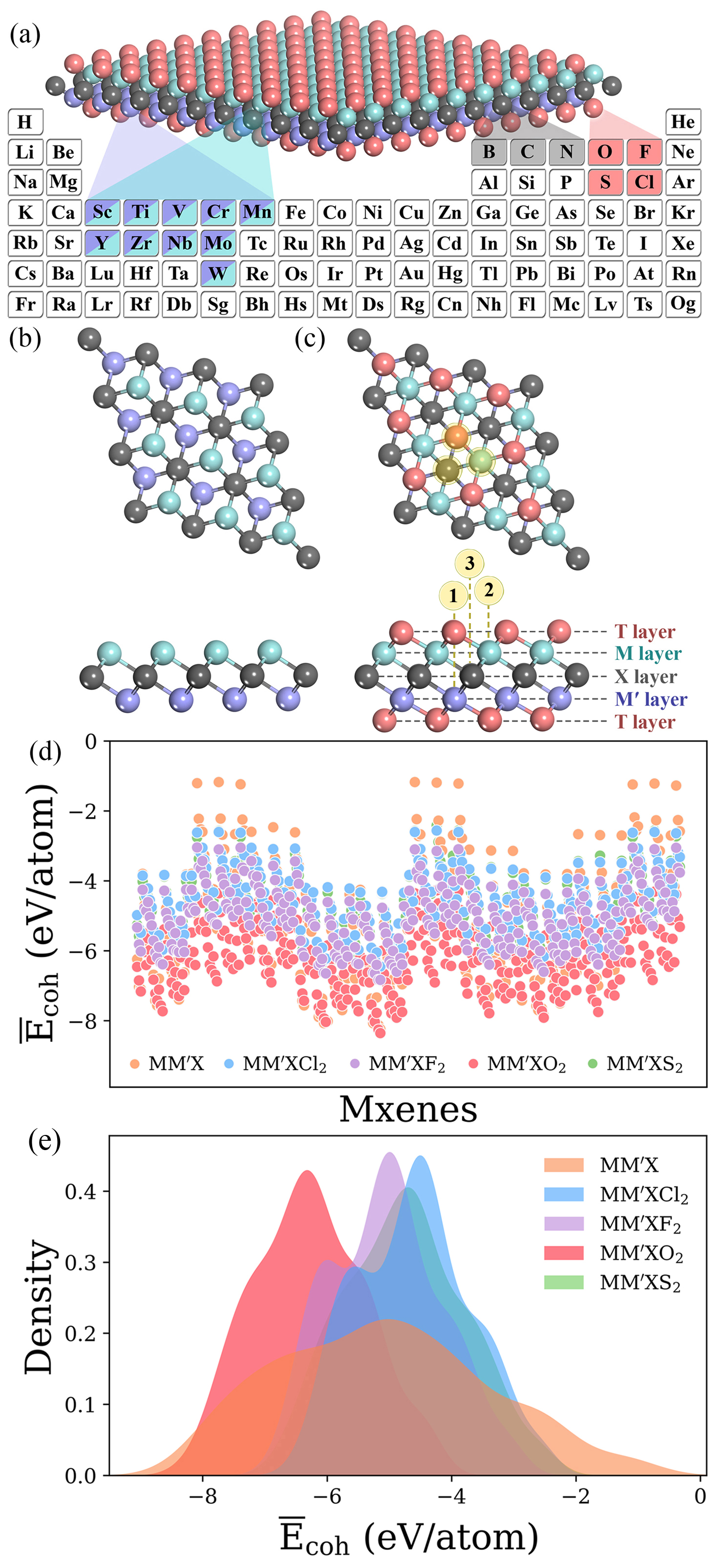}}}
\caption{(a) Selected elements for MM$^{\prime}$XT$_2$ MXenes (M/M$^{\prime}$ = Sc, Ti, V, Cr, Mn, Y, Zr, Nb, Mo or W; X = B, C or N; T = O, F, Cl or S). Optimized structures of (b) pristine and (c) functionalized MXenes.  Color code: M/M$^{\prime}$-, X- and T- layers are presented in blue/voilet, dark grey and pink colors, respectively. The number 1, 2 and 3 in circle indicates possible adsorption sites. (d) Computed normalized cohesive energies $\overline{E}_{coh}$ (eV/atom) and (e) corresponding distribution of MM$^{\prime}$X, MM$^{\prime}$XCl$_2$, MM$^{\prime}$XF$_2$, MM$^{\prime}$XO$_2$ and MM$^{\prime}$XS$_2$ MXenes.  }
\end{figure}
 
 number of MXene configurations.

\section{Results}
MM$^{\prime}$XT$_2$-type (M/M$^{\prime}$ = Sc, Ti, V, Cr, Mn, Y, Zr, Nb, Mo or W; X = B, C or N; T = O, F, Cl or S) MXenes were constructed through quintuple atomic layers of T-M-X-M$^{\prime}$-T, where the X layers are alternately sandwiched between different metal layers (M/M$^{\prime}$) and the surfaces are terminated with functional groups (T) as shown in Fig. 2a-c. Possible combinations of metal layers were then considered to generate 1,500 MM$^{\prime}$XT$_2$-type MXenes. Initially, we evaluated the cohesive energies to understand the stability trends in these configurations. Computed normalized cohesive energies $\overline{E}_{coh}$ (eV/atom) and the corresponding distribution of various functionalized MXenes are shown in Fig. 2d,e. From the viewpoint of functionalization, the lowest $\overline{E}_{coh}$ is obtained for -O terminated MXenes when compared with other terminations such as -F, -Cl and -S, which indicates that the former is more likely to be synthesized during experimentation. Moreover, the structural stability of the terminated MXenes increases in the order of MM$^{\prime}$X $<$ MM$^{\prime}$XCl$_2$ $<$ MM$^{\prime}$XS$_2$ $<$ MM$^{\prime}$XF$_2$ $<$ MM$^{\prime}$XO$_2$, representing better stability for fully functionalized MXenes with respect to their pristine counterpart. The observed behavior also confirms why the MXenes are usually terminated with functional groups during experimental synthesis\cite{NaguibMM}. In addition, the $\overline{E}_{coh}$ values of MM$^{\prime}$CT$_2$ are lower than those of MM$^{\prime}$BT$_2$ and MM$^{\prime}$NT$_2$, suggesting that the carbon based MXenes are more stable than the boride and nitride based MXenes (see Fig S1a,b). This also provides an alternative explanation for the poor stability in the etching of boride and nitride based MXenes during the synthesis process\cite{Soundir,NgVMH}.

\subsection{Adsorption energy distribution}

Typically, the  availability of active catalytic sites on the surface of MXenes is highly required to carry out the HER activity: the larger the number of active sites on the surface, the stronger will be the catalytic performance. There are three possible adsorption sites available on MXene surfaces for hydrogen adsorption. Site-1 represents that the H-atom is adsorbed directly on the innermost metal atomic layer of the MXenes, site-2 indicates that the H-atom is adsorbed on the top of the outermost M-atom of MXenes and site-3 denotes that the H-atom is adsorbed directly above the X-atom of MXene  structures. Overall, the adsorption of H-atom on the three available active sites of 1,500 MM$^{\prime}$XT$_2$-type MXenes leads to 4,500 configurations. Among them, 1,125 systems (25\% of the materials' space) were randomly selected to evaluate the HER activity using DFT calculations as well as for training the ML models; while the catalytic performance for the rest of the materials was predicted using the well-trained ML model. Based on the computational hydrogen electrode (CHE) model\cite{JGreeley}, the Gibbs free energy of adsorbed hydrogen ($\Delta$G$_{H}$) is  a universal indicator to evaluate the HER performance.

\begin{figure*}[ht]
\centering
{{\includegraphics[height = 4.2in,width=7in]{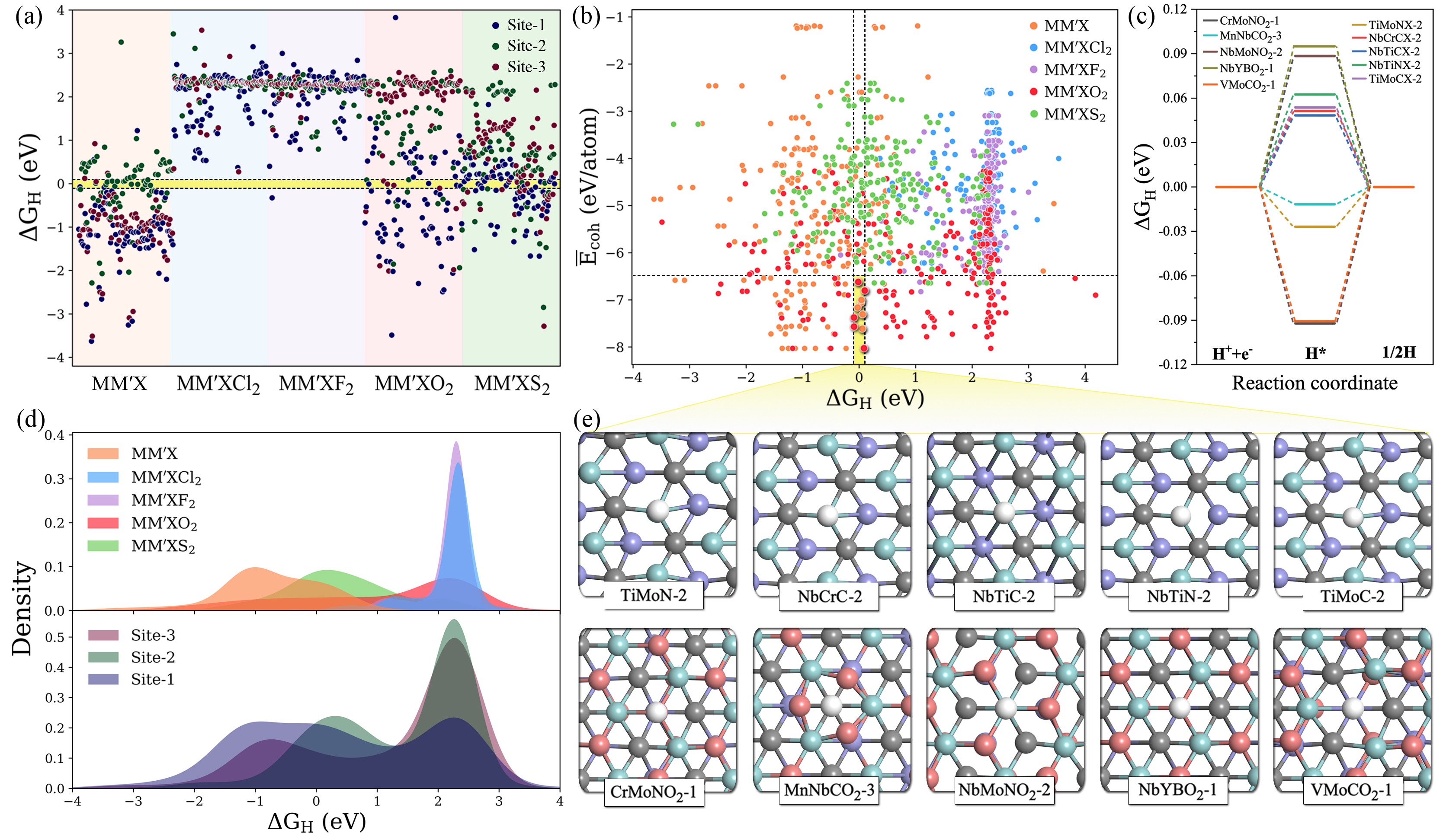}}}
\caption{(a) DFT computed hydrogen adsorbed Gibbs free energies ($\Delta$G$_{H}$) for randomly selected 1,125 MM$^{\prime}$XT$_2$-type MXenes. $\Delta$G$_{H}$ in the range of -0.1 to +0.1 eV is represented in the yellow shadow region. (b) Normalized cohesive energies $\overline{E}_{coh}$ versus $\Delta$G$_{H}$. The top 10 promising candidates with better stability and high HER activity are highlighted in the yellow region. (c) The free energy profile of hydrogen evolution for the top 10 potential candidates. (d) Distribution of $\Delta$G$_{H}$ with respect to functionalization and type of active sites. (e) Optimized geometries of hydrogen adsorbed on the top 10 promising MXenes. Here blue, violet, grey, pink and white color balls represent M, M$^{\prime}$, X and T, H atoms, respectively. }
\end{figure*}

 Accordingly, the $\mid\Delta G_{H}\mid$ close to zero signifies prominent HER activity of the catalyst; while, a negative or positive $\Delta$G$_{H}$ with too strong or too weak adsorption will tend to reduce the overall reaction rate.  The HER performance is highly dependent on functionalization (see Fig. 3a-e and S2); for instance, most of the F- and Cl- terminated MXenes exhibit poor HER activity due to their highly positive $\Delta$G$_{H}$, indicating a weak interaction between adsorbed H, and F- and Cl- groups on the MXene surfaces. There is a significant difference in HER activity even with varying X-layers, where the carbon based MXenes with $\mid\Delta$G$_{H}\mid$ smaller than 0.1 eV show better HER performance when compared to boride and nitride based MXenes. Overall, 48 systems show optimal $\Delta$G$_{H}$ values in the range of -0.1 to 0.1 eV. Among them, CrMoNO$_2$-1, MnNbCO$_2$-3, NbMoNO$_2$-3, NbYBO$_2$-1, VMoCO$_2$-1, TiMoN-2, NbCrC-2, NbTiC-2, NbTiN-2 and TiMoC-2 have better stability and superior HER activities when compared with the noble metal Pt\cite{ObodoKO} and thus can be considered as promising HER catalysts. These results reveal that the HER activity also depends on the active site where the H is adsorbed. It is found that the H adsorbed directly on the outermost metal atomic layer of the MXene structures (site-2) has better HER catalytic performance when compared with other sites.

\begin{figure*}[t]
\centering
{{\includegraphics[height = 3.3in,width=7in]{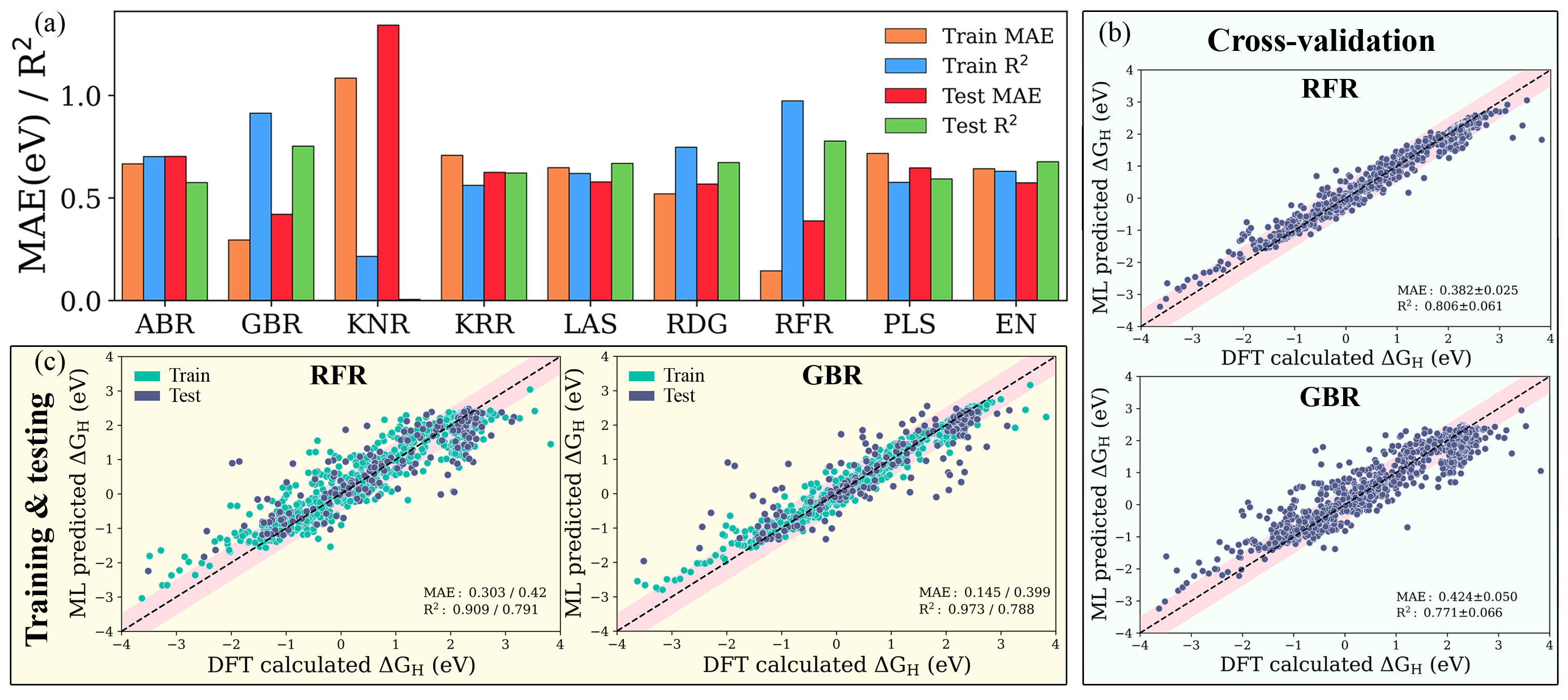}}}
\caption{Mean absolute error (MAE)and coefficient of determination (R$^2$ score) of  ABR, ENR, GBR, KNR, KRR, LAS, PLS, RFR and RDG algorithms using primary (atomistic, structural and electronic indicators) and statistical function processed features. Parity plot of the best-performing RFR and GBR models (b) with and (c) without cross-validation using the DFT dataset of hydrogen adsorbed Gibbs free energies ($\Delta$G$_{H}$). The pink-shaded region indicates a deviation of up to 0.5 eV.}
\end{figure*}

\subsection{ML models optimization}
The precision of a well-trained ML model mainly depends on the material descriptor as well as the choice of algorithm. Historically, several aspects have been considered to connect with the chemical reactivity of catalytic materials, such as d-band characteristics, coordination number and bulk or atomic properties. Correlating such physical aspects onto the adsorption energies with highly non-linear regression algorithms even requires features from the fully optimized geometries of the clean adsorption sites. With the subset selection of atomistic, surface and statistical features, we establish nine different ML models as shown in Table S3, namely, ABR, ENR, GBR, KNR, KRR, LAS, PLS, RFR and RDG using the dataset containing 1,125 H adsorption energies obtained from DFT calculations. To ensure the accuracy and generalization of the supervised ML models, we partitioned the data into training and test sets in an 80 by 20 ratio (see Fig. S3). For controlling and assessing against overfitting, the coefficient of determination value (R$^2$ score) and mean absolute error (MAE) were estimated with and without using 10-fold cross-validation technique. As shown in Fig. S4 and Table S4, the subset of features with representative physical indicators anisotropically captures the H adsorption energy over the studied MXenes. Among all the subsets of features, the combination of primary features with the indicators processed through statistical functions provides the best predictive performance. The predicted MAE and R$^2$ of  ABR, ENR, GBR, KNR, KRR, LAS, PLS, RFR and RDG algorithms using primary and statistical function processed features are presented in Fig. 4a. The use of the RFR model converges to low MAE with highest R$^2$ score, irrespective of the feature subset and thereby demonstrating its good generalization ability. Predictions by the best-performing RFR and GBR models with and without cross-validation using the DFT dataset of hydrogen adsorbed Gibbs free energies ($\Delta$G$_{H}$) are shown in Fig. 4b and Fig. 4c, respectively. The GBR model also shows better performance with R$^2$ score of 0.913 (0.753) and MAE of 0.294 (0.421) eV in the model training (testing), indicating the inferior accuracy prediction to the RFR model. It should be noted that these tree based RFR and GBR ensemble models are robust against high-dimensional data sets due to the high ability to fit nonlinear data. On the other hand, ABR, ENR,  KNR, KRR, LAS, PLS and RDG methods have unsatisfactory prediction performance, which is reflected by their considerable MAEs of 0.702, 0.573, 1.342, 0.625, 0.578, 0.647, 0.568 eV (see Table S5), respectively, due to poor extrapolation capabilities of the models. Using 10-fold cross validation, the studied models exhibit similar  prediction performance for the training/testing sets as shown in Fig. S5. These results demonstrate that the materials' descriptors are crucial to reproduce the adsorption energies over MM$^{\prime}$XT$_2$-type MXenes, thereby validating the suitability of our feature pool. The combination of primary and statistical features achieved satisfactory prediction accuracy. Nevertheless, the presence of a large number of input features makes it difficult to readily derive physical insights, thereby increasing the complexity and time-consumption in the ML model. Thus, it is important to look for a fine balance between accuracy and the number of features for obtaining efficient results.

\begin{figure*}[t]
\centering
{{\includegraphics[height = 4.5in,width=7in]{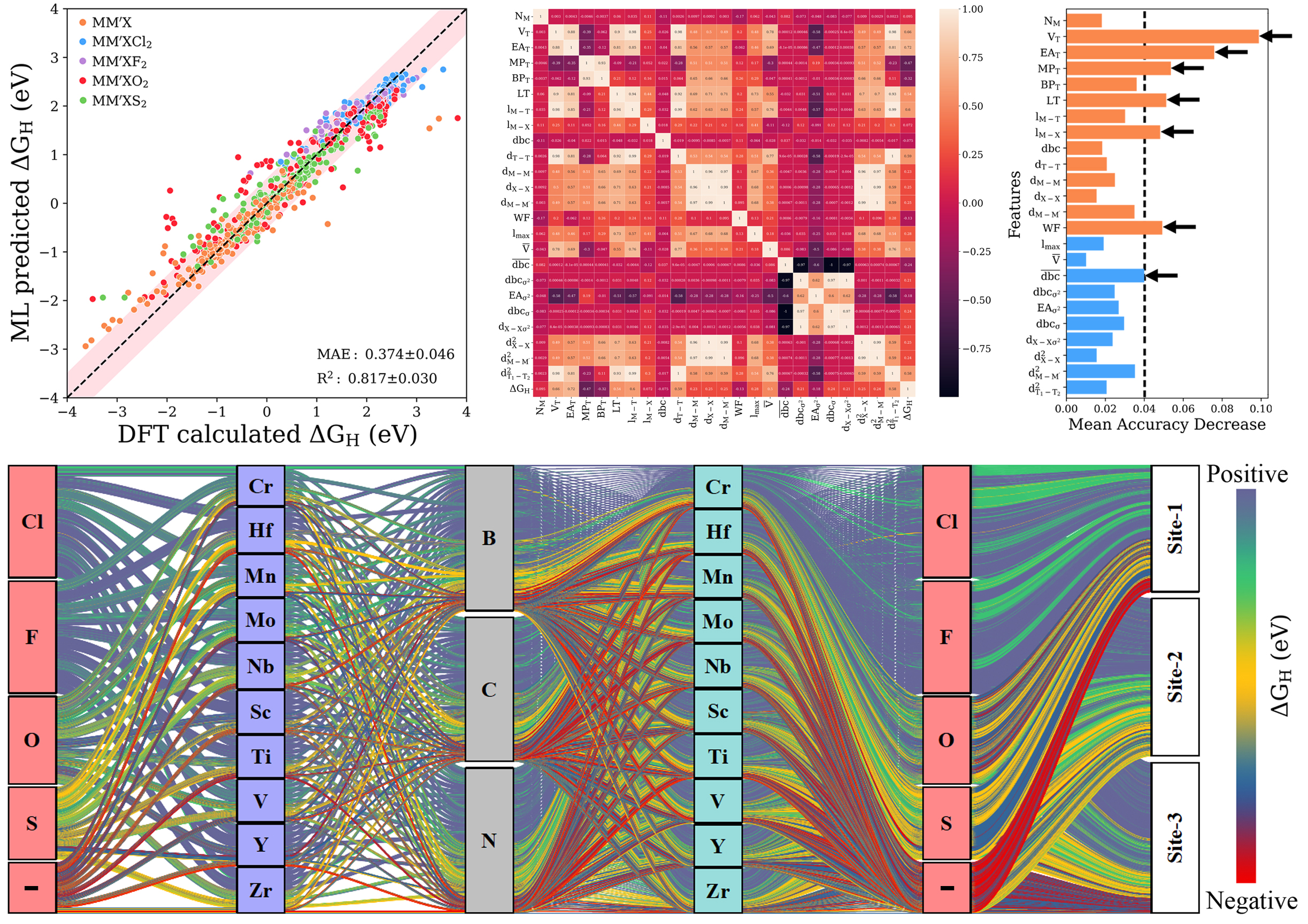}}}
\caption{(a) Parity plot of predicted vs actual $\Delta$G$_{H}$ by RFR model with RFE-HO in the best cross-validated process. The pink-shaded region indicates a deviation of up to 0.5 eV. (b) Pearson correlation coefficient (PCC) heat map for the reduced set of features after recursive feature elimination (RFE) and  hyperparameter optimization (HO), (c) Feature importance using permutation on the RFR model with RFE-HO, evaluated via 10-fold cross-validation. (d) Alluvial diagram for the predicted $\Delta$G$_{H}$ values of 4,500 MM$^{\prime}$XT$_2$-type MXenes. The positive, negative and optimal $\Delta$G$_{H}$ values are represented in blue, red and yellow colors. The "-" symbol indicates pristine MXenes without termination. Clearly, the Cl- and F- functionalizations show blue color links indicating poor HER activity due to highly positive $\Delta$G$_{H}$; while H adsorbed directly on the outermost metal atomic layer of the MXene structures (site-2) have better HER catalytic performance as shown by yellow color links. }
\end{figure*}

\subsection{Recursive feature elimination and  hyperparameter optimization}
Identifying the most representative descriptors is an extremely critical step for feature engineering to minimize the prediction biasing and accelerate the efficiency of the ML model. For this purpose, recursive feature elimination (RFE) is used to filter the descriptors with extreme asymmetry (skewness) and with low/zero variance for recognizing more suitable smaller subset of features. In addition, hyperparameter optimization (HO) was performed on RFR and GBR models by varying the range of parameters using 10-fold cross-validation and the best combinations of hyperparameters were presented in Table S6. RFE decreased the number of features from 125 to 24 and 30 for RFR and GBR, respectively. Clearly, the reduced set of features is found to be sufficient for capturing the complex interactions influencing the Gibbs free energies. The performance of RFR and GBR models show slight improvement through RFE-HO, in both efficiency and predicted accuracy. The MAE/R$^2$  values of RFR and GBR models with RFE-HO are 0.37/0.82 and 0.37/0.81 (see Fig. 5a, S6-S8 and Table S7), respectively, demonstrating that the RFR model is slightly more suitable and the best algorithm in our multistep ML workflow. In addition, the high ranking of a descriptor  indicates the vital role in governing the HER activity of MXenes; for example, the valance electron number of termination (V$_T$) predominately affects the adsorption ability of H atom. As the typical descriptor sets generated from RFE-HO process vary with RFR and GBR models, we particularly identified common potential descriptors that precisely connect to the physicochemical properties of MXenes. Subsequently, valence electron number (V$_T$) and electron affinity  (EA$_T$) of termination and work function (WF) are the strong key predictors of Gibbs free energy. As shown in Fig. 5b, the Pearson correlation coefficient (PCC) heat map for the reduced set of features is basically consistent with the ranking of features. However, there exists a strong correlations among some statistical and primary features. This is because the statistical features are derived from the same primary features. For instance, the distance between the metal atoms (d$_{M-M}$), functionalized atoms (d$_{T-T}$), X-atoms (d$_{X-X}$) is strongly correlated to the square of their corresponding distances. Overall, the mean accuracy decrease of the model indicates key role of the down-selected descriptors in determining the HER activity (see Fig. 5c, S9 and Table S8, S9), where the further removal of any feature from the list may lead to a relative decrease in the efficacy of the model.

\subsection{Performance prediction of the unknown space}

After developing the well-trained model, the best-performing RFR and GBR strategies with RFE-HO were further applied to the remaining 3,375 MM$^{\prime}$XT$_2$-type MXenes. As mentioned, the well-trained RFR model with RFE-HO through cross-validation has an MAE of 0.37 for the randomly selected 25\% of the materials' space (1,125 systems). Thus, the optimal $\Delta$G$_{H}$ value for the remaining ML predicted materials' space were set as -0.470 (-0.1 - 0.370 eV) $\sim$ 0.470 (0.1 + 0.370 eV). Fig. 5d presents the $\Delta$G$_{H}$ for the complete list of MM$^{\prime}$XT$_2$-type MXenes using well-trained ML methodology. Clearly, the $\Delta$G$_{H}$ is anisotropically distributed over a large energy scale ranging from 2.75 to -2.94 eV, indicating the substantial heterogeneity of the active sites.  Out of 4,500 MM$^{\prime}$XT$_2$-type MXenes, 28 candidates show optimal $\Delta$G$_{H}$ which signifies excellent HER catalyst activity (see Table S10). Similar to the DFT computed results, the thermodynamic uphill in the proton adsorption energies on F- and Cl- functionalized MXenes suppress the hydrogen production activity, while O- and S- termination show better optimal $\Delta$G$_{H}$ values. These results reveal that the HER activity mainly depends on the type of functionalization. In addition, the type of active sites also influences the catalytic activity, where site-2 is found to exhibit efficient HER performance (see Fig. S10a-c). In a broader context, our calculation results indicate that the H adsorbed on site-2 with some specific elements, such as O-functionalization of C-layer can be used to make MM$^{\prime}$XT$_2$-type MXenes suitable for enhancing the HER activity.

\section{Discussion}
We have developed a multistep workflow for rapid and accurate $\Delta$G$_{H}$ predictions of 4,500 MM$^{\prime}$XT$_2$-type MXenes, in which 1,125 systems were randomly selected as the training samples to evaluate the HER performance using DFT calculations. These MXenes show high structural stability, especially -O terminated structures are highly preferable and  more likely to be synthesized during experimentation. From DFT computed $\Delta$G$_{H}$, we noticed weak interaction between hydrogen and F- and Cl- functional groups of MXenes, indicating poor HER activity. While, O- and S- terminations show better HER performance due to moderate hydrogen adsorption with $\Delta$G$_{H}$ close to zero. It should be noted that the carbon based MXenes are more preferable when compared to boron and nitride based structures due to their better HER activity. Our results demonstrate that the hydrogen adsorbed directly on the outermost metal atomic layer of the MXene structures (site-2) is beneficial to enhance the HER performance. During ML model optimization, the primary features are unable to capture the trends of the target property and thus we have considered the statistical measures of the indicators, where the number of features was increased from 60 to 125. The RFR and GBR models show better performance when compared with the other studied algorithms.  Subsequently, the recursive feature elimination (RFE) method is employed to filter the features  with low/zero variance and with extreme asymmetry (skewness) for recognizing more suitable smaller subset of descriptors. The RFR model with RFE-HO found to exibit best predictive performance towards $\Delta$G$_{H}$ with low MAE and high R$^2$ of 0.37 eV and 0.82, respectively. These results demonstrate that the materials descriptors are crucial to reproduce the adsorption energies over MM$^{\prime}$XT$_2$-type MXenes, thereby validating the suitability of our feature pool. The feature importance analysis revealed valence electron number (V$_T$) and electron affinity (EA$_T$) of termination, and work function (WF) as key descriptors that govern the HER performance. The final RFR model is then used to predict the HER activity of the remaining materials' space. We found that the C-layers alternately sandwiched between  Nb, V, Mo, Cr, Ti metal layers of O-functionalized MM$^{\prime}$XT$_2$-type MXenes show high stability and better HER activity. In conclusion, the present work not only established a robust and more broadly applicable ML-DFT based multistep workflow for efficient and accurate screening of HER activity but also provides potential factors that govern the efficiency of the catalysts, thereby accelerating the design and development of novel catalysts with high performance.

\section{Methods}
\subsection{Density Functional Theory}
Ab initio simulations were performed using plane-wave based Vienna ab initio simulation package (VASP) code within the framework of density functional theory. The exchange correlation effects and ion-electron interactions were incorporated through GGA-functional by Perdue-Burke-Ernzerhof and Projected Augmented Wave method (PAW), respectively. For structural relaxation, the convergence thresholds of 10$^{-2}$ eV/\AA \space in force and 10$^{-5}$ eV in energy, with cutoff energy and Monkhorst-Pack method  k-point grid of 450 eV and 7 $\times$ 7 $\times$ 1, respectively, were employed to expand the electron wave functions and sampling the Brillouin zone. A vacuum space of 15 \AA \space was adopted along the z-direction to prevent spurious interaction among the periodic units. The Grimme's empirical correction scheme (DFT + D3) was adopted to describe the Van der Waals interactions.

The hydrogen adsorbed Gibbs free energy ($\Delta$G$_{H}$) was defined based on the computational hydrogen electrode (CHE) model\cite{NorskovJK} as shown below:
\begin{equation}
\Delta G_{H} = \Delta E_{H} + \Delta E_{ZPE} - T\Delta S
\end{equation}
where $\Delta$E$_{H}$ is the DFT computed differential hydrogen adsorption energy. $\Delta$E$_{ZPE}$, T and $\Delta$S are the change in the zero-point energy of each term contribution, temperature (298.15 K) and entropy change, respectively, calculated in the harmonic approximation. $\Delta$E$_{H}$ can be calculated as follows:
\begin{equation}
 \Delta E_{H} =  E_{H} - E_{slab} - \frac{1}{2}E_{H_2}
\end{equation}
where E$_{slab}$, E$_{H}$  and E$_{H_2}$ are the total energies before H-adsorption, after H-adsorption and isolated H$_2$ gas molecule, respectively. Attending to this
definition of $\Delta$G$_{H}$, the highly positive or highly negative values are detrimental as they act as a large barrier to the electrochemical reduction reaction and make it difficult during H$_2$ desorption. While the optimal $\Delta$G$_{H}$ values close to zero are highly preferable to obtain an excellent HER catalyst.

The cohesive energy (E$_{coh}$) which is the measure of total energy of the system abstracted by the sum of the individual constituent atoms energies can be used to determine the structural stability by understanding the strength of forces that binds the atoms together in a system and is defined as follows:

\begin{equation}
 E_{coh} = E_{Total} - NE_M - NE_{M^{\prime}} - NE_X - NE_T
\end{equation}
where E$_{Total}$ is the total energy of the system. E$_{M}$/E$_{M^{\prime}}$, E$_X$ and E$_T$ are the energies of free atoms of M (M = Sc, Ti, V, Cr, Mn, Y, Zr, Nb, Mo or W), X (X = B, C or N) and T (T = O, F, S or Cl), respectively. We further computed the cohesive energy per atom by normalizing the E$_{coh}$ of different systems: $\overline{E}_{coh}$ = $\frac{E_{coh}}{No. \space of atoms}$.

\subsection{Feature space construction}
To establish accurate ML models or to evaluate the main contributions ruling the hydrogen evolution reaction, it is important to map the material-to-attribute connection. On this account, a group of features (material variables) that represent a system in a computer-friendly manner is highly required. Typically, an ideal feature set discloses the structure-activity relationship of a system and specifically describes each material input data set. However, the materials' representation is a concern of complex and intense development, where the explicit interpretation displays a whopping challenge when compared with the success recently attained for molecular representations\cite{Pronobis,Faber}.  Thus, it is very important to generate suitable and comprehensive features during the construction of ML models. For easy accessibility of training the efficient and fast ML model, every selected feature has to independently represent the physicochemical property. Within this purpose, we have considered atomistic, structural and electronic indicators as an initial pool of descriptors, which leads to a total number of 60 primary features as shown in Table S1.  Nevertheless, the selected primary features were unable to capture the HER performance due to different numbers of constituent atoms that have different feature space sizes. To this end, the statistical measures of some selected primary features were considered, including average, weighted average, maximum, minimum, standard deviation, variance, and squared values (see Table S2). Feature addition using statistical functions tallied features to 125. These features are categorized into Set-1 (atomistic features), Set-2 (surface features) and Set-3 (statistical features) and their corresponding subset combinations are employed to identify the key descriptors. Albeit, the considered descriptors may not provide complete information about the fundamental physicochemical principles. However, in a pragmatic outlook, their predictions can be used as an indicator to understand the importance of variables that influence the property of interest, thereby establishing a potential and practical model to replace the complicated problem.

\subsection{Machine Learning}
Our ML approach is designed for establishing a regression relationship between the  HER catalytic activity and predominating indicators, based on the results from DFT calculations. Nine ML algorithms, namely, AdaBoost Regressor (ABR), Elastic Net Regressor (ENR), Gradient Boosting Regressor (GBR), K Neighbors Regressor (KNR), Kernel Ridge regressor (KRR), Lasso (LAS), Partial Least Squares (PLS), Random Forest Regressor (RFR) and Ridge Regression (RDG) were employed to predict the HER performance. An open-source Python distribution platform is used to train the models via scikit-learn libraries. The 25\% of the materials' space (1,125 systems) is randomly selected to evaluate the HER performance using density functional theory (DFT) calculations, while the activity of the remaining 85\% materials' space is predicted through the well-trained ML model. To ensure the accuracy and generalization of the supervised ML models, The H binding energy data obtained from DFT-calculations were randomly partitioned into training and test sets in an 80 by 20 ratio. The stability and accuracy of all models were evaluated through the coefficient of determination (R$^2$) and mean absolute error (MAE), with standard deviations indicated, and formulas are given as:

\begin{equation}
R^2 = 1- \frac{ \sum_{i=1}^{n} (y_i - \overline{y})^2}{ \sum_{i=1}^{n} (\dot{y}_i - \overline{y})^2}
\end{equation}

 \vspace{0.3cm}

\begin{equation}
MAE = \frac{\sum_{i=1}^{n} (y_i - \dot{y}_i)^2}{n}
\end{equation}
where ${y_i}$, $\dot{y}$, and $\overline{y}$ are the prediction, true and average value, respectively.  R$^2$ score ranges from 0 to 1. The model with R$^2$ value (MAE) closer to 1 (0) demonstrates the better model performance.

\vspace{0.2in}
{\large\textbf{Acknowledgements}\\}
This work is supported by the Department of Science and Technology, Government of India, under the grant number SPO/DST/CHE/2021535. BMA would like to thank SERB for the financial support under the grant number PDF/2021/000487. We acknowledge National Supercomputing Mission (NSM) for providing computing resources of PARAM Sanganak at IIT Kanpur.

\subsection{Data availability}
The dataset consisting all the features along with the DFT calculated ${{\Delta} G_H}$ values of 1,125 MXenes are available on our GitHub repository. The primary features' data was collected from chemical repository. The best RFR and GBR models can be accessed from zip folder to further predict ${{\Delta} G_H}$ of MM$^{\prime}$XT$_2$-type MXenes, provided the aforementioned descriptors.

\subsection{Code availability}
The code can be retrieved from our GitHub repository (https://github.com/cnislab/MXenes4HER) and Zenodo (https://doi.org/10.5281/zenodo.7414537). Several python libraries are employed in the current work: Pandas to analyse the dataset, SciKit-learn to build the regression models, Joblib to save the best models, and Matplotlib together with Seaborn to visualize the plots.

\end{document}